\documentclass[%
 Revtex4, 
 amsmath,amssymb,
 aps, physrev,
]{revtex4-2}

\usepackage{graphicx}
\usepackage{dcolumn}
\usepackage{xcolor}
\usepackage{ulem}
\usepackage{bm}
\usepackage{float,soul,caption,mathtools}
\usepackage[colorlinks=true,linkcolor=blue,citecolor=blue,urlcolor=blue]{hyperref}

\usepackage{amsmath, amssymb}
\usepackage{pgfplots}
\usepackage{subcaption}

\pgfplotsset{
    compat=1.18,
    label style={font=\small},
    tick label style={font=\small},
    legend style={font=\small},
    every axis/.append style={
        line width=0.8pt,
        tick style={line width=0.6pt, color=black},
        grid style={line width=0.4pt, color=gray!30},
        clip=false
    }
}

\definecolor{softred}{HTML}{E41A1C}   
\definecolor{softblue}{HTML}{377EB8}    
\definecolor{lightred}{HTML}{FFCCCC}    

\begin{document}

\preprint{....}

\title{\textbf{Analytic structure of the QCD phase diagram in the complex-temperature plane} 
}%

\author{G\"ok\c ce Ba\c sar}
 \email{gbasar@unc.edu}
 \affiliation{
Department of Physics and Astronomy,
University of North Carolina at Chapel Hill, Chapel Hill, NC, 27599
 }

\author{Vladimir V. Skokov}%
 \email{vskokov@ncsu.edu}
\affiliation{%
Department of Physics and Astronomy,
North Carolina State University,
Raleigh, NC 27695
}%

\date{\today}

\begin{abstract}
We study the analytic structure of the QCD phase diagram by treating temperature as a complex variable. The nearest Yang--Lee edge singularities in the complex $T$ plane bound the domain of analyticity of temperature-dependent thermodynamic observables and complement the more commonly studied singularities in the complex chemical-potential plane. Our analysis combines three complementary perspectives: universal critical scaling, a first-principles extraction from lattice-QCD data, and explicit illustrations in effective models. We illustrate the resulting structure in a random-matrix model and in a quark--meson model, where the singularity trajectories can be followed explicitly. At small real chemical potential, the leading complex-temperature singularity admits an analytic expansion in $\mu^2$, while near a critical point it crosses over to the universal Puiseux form dictated by Ising critical scaling. We show that the complex-$T$ and complex-$\mu$ trajectories are controlled by the same scaling variables and mapping coefficients, so their comparison provides a stringent consistency test of critical-point searches and constrains the extent of the critical scaling regime. Finally, we analyze lattice-QCD data at $\mu=0$ using an iterated conformal--Pad\'e approach and extract the continuum location of the nearest complex-temperature singularity. The result is consistent with the expectation that, at physical quark masses, the real part of the leading singularity lies between the chiral-limit transition temperature and the physical-mass chiral-susceptibility peak temperature, while its imaginary part remains nonzero.
%
\end{abstract}

\maketitle


\section{\label{sec:intro} Introduction}

The thermodynamic functions of QCD are analytic only within domains bounded by singularities in the complexified control parameters.  Although phase transitions occur on the real axes only in special limits, nearby complex singularities leave {characteristic signatures in} Taylor expansions, Pad\'e approximants, Fourier coefficients, and analytic continuations from imaginary chemical potential or from real-temperature data.  In this sense, the analytic structure of the QCD partition function provides a quantitative bridge between finite-volume Lee--Yang zeros, universal critical scaling, and the practical problem of locating the QCD critical point~\cite{Stephanov:2006dn,An:2016lni,An:2017brc,Mukherjee:2019eou,Connelly:2020gwa,Rennecke:2022ohx,Johnson:2022cqv,Basar:2023nkp,Karsch:2023rfb,Skokov:2024fac,Wada:2024qsk,Bryant:2024twd,Clarke:2024ugt,Wan:2024xeu,Adam:2025phc,Wan:2025wdg,Wada:2025ghc,Wada:2026qzt,Basar:2026irk}.

Most existing discussions focus on the complex chemical-potential plane.  This is natural for finite-density QCD, because Taylor expansions in $\mu/T$ and simulations at imaginary chemical potential both probe the analytic continuation in $\mu$.  However, the temperature is an equally legitimate complex variable.  The nearest singularities in the complex $T$ plane determine the analytic domain of thermodynamic observables as functions of temperature, and they can be accessed directly from high-precision lattice data at $\mu=0$.  Moreover, if the singularities are controlled by the same critical scaling fields, their trajectories in the complex $T$ and complex $\mu$ planes cannot be independent.  Establishing this relation is the main goal of this work.

We use the term Yang--Lee edge (YLE) for the branch-point singularities obtained after {analytic continuation of the thermodynamic control parameters into the complex plane}. {Strictly speaking, near a critical point,  a singularity reached by complexifying temperature or chemical potential in QCD are neither YLE nor Fisher-like; throughout this work, however, we use YLE as a common shorthand for the edge singularity associated with phase transitions.}  At nonzero quark mass{,} the chiral transition is a crossover on the real axis, but the remnant of the chiral critical singularity survives as a pair of complex-conjugate YLE singularities, analogous to the pair in the complex chemical-potential plane~\cite{Stephanov:2006dn}.  Near a critical point, the location of the YLE is fixed by the universal scaling equation of state together with a nonuniversal map from QCD variables to Ising variables.  Consequently, the motion of the singularity with real $\mu$ in the complex $T$ plane and its motion with real $T$ in the complex $\mu$ plane are controlled by the same set of parameters.  This shared structure provides a practical test: a singularity extracted in one complex plane should predict the leading trajectory in the other complex plane if both are governed by the same critical scaling regime.

The paper is organized as follows.  Section~\ref{sec:rmm} uses a random-matrix model to introduce complex-temperature singularities and to compare the small-$\mu$ expansion with the critical scaling form near the critical point.  Section~\ref{sec:qm} repeats the analysis in a quark--meson model, where changing the vacuum scale changes the critical-line geometry and the approach of the YLE to the critical point.  Section~\ref{sec:critical-scaling} derives the scaling relations that connect YLE trajectories in the complex $T$ and complex $\mu$ planes.  Section~\ref{sec:lattice} describes the extraction of the nearest complex-temperature singularity from lattice-QCD data at $\mu=0$.  Section~\ref{sec:conclusions} summarizes the results.  Technical details of the thermodynamic potential and a special critical-point mapping are collected in the appendices.

\section{Illustrations in mean-field models}

\subsection{\label{sec:rmm} Random matrix model}
The chiral random-matrix model (RMM)~\cite{Halasz:1998qr} is a toy model of QCD that captures chiral symmetry breaking and contains a critical point on the line separating the crossover and first-order regimes.  Its simplicity also allows the relevant branch-point singularities to be tracked explicitly. In this model, the matrix elements of the Dirac operator are replaced by Gaussian random variables.
In the limit where the number of eigenvalues $N\rightarrow \infty$, the model is described by a single order-parameter variable $\phi$ with the effective potential
\begin{equation}
    \Omega(\phi;T,\mu,m)=A\,N_f \left(\phi^2-\frac12 \log\left\{\left[(\phi+m)^2-(\mu+iT)^2)\right]\cdot\left[(\phi+m)^2-(\mu-iT)^2\right]\right\} \right)
\end{equation}
where $m$ is the quark mass, $N_f$ is the number of fermion species and $A$ is a phenomenological constant. All dimensionful variables $T$, $\mu$, $\phi$, and $m$ are expressed in units of the chiral transition temperature at zero chemical potential.  

In the large $N$ limit, the equilibrium properties of the model are determined by minimizing the grand potential. In particular, the pressure is obtained by evaluating the potential at the saddle point,
\begin{equation}
    p(T,\mu,m)=-\Omega(\phi^*(T,\mu,m);T,\mu,m),
\end{equation}
where
\begin{equation}
   \left. \frac{\partial}{\partial\phi}\Omega(\phi;T,\mu,m)\right|_{\phi=\phi^*}=0\,.
\end{equation}
The condensate and chiral susceptibility are then 
\begin{equation}
\bar\psi\psi=\phi^*(T,\mu,m), \quad \chi=\frac{\partial \phi^*}{\partial m}\,.
\end{equation}

\begin{figure}[ht]
    \centering
    \includegraphics[width=0.39\linewidth]{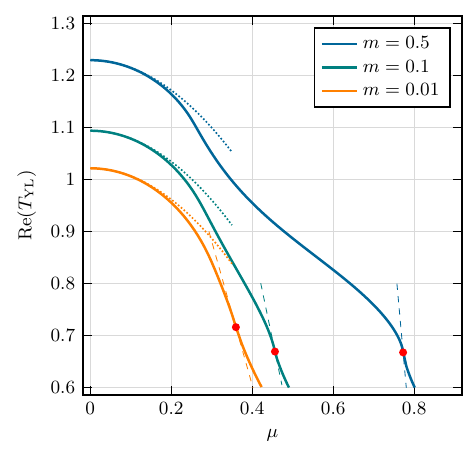}
    \includegraphics[width=0.39\linewidth]{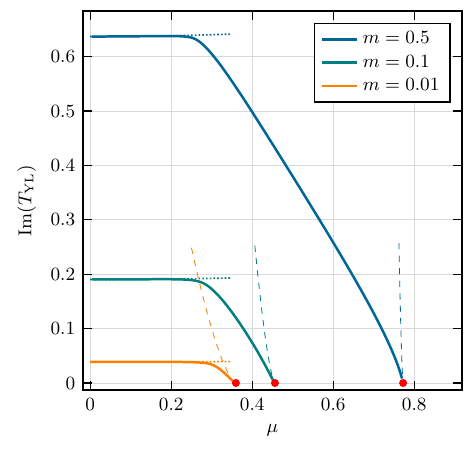}
    \caption{Real and imaginary parts of the leading Yang--Lee edge singularity in the complex $T$ plane as functions of the real chemical potential in the RMM.  The dotted curves show the small-$\mu$ expansion, while the dashed curves show the critical scaling form near the critical point, marked by red points.}
    \label{fig:T-rmm}
\end{figure}

At a YLE singularity, the susceptibility diverges, i.e. $\chi^{-1} = 0$, which is equivalent to 
\begin{equation}
   \left. \frac{\partial^2}{\partial\phi^2}\Omega(\phi;T,\mu,m)\right|_{\phi=\phi^*, T=T_{\rm YLE}}=0\,.
\end{equation}
This equation implicitly determines $T_{\rm YLE}$ as a complex function of the real parameter $\mu$.  The real and imaginary parts of this function for different values of $m$ are shown in Fig.~\ref{fig:T-rmm}. 

\begin{figure}
    \centering
\includegraphics[width=0.36\linewidth]{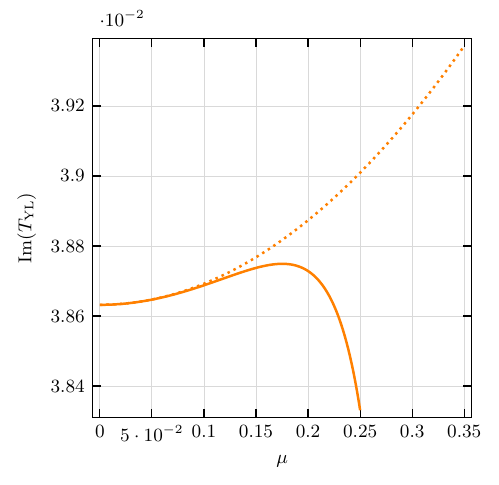}
  \includegraphics[width=0.39\linewidth]{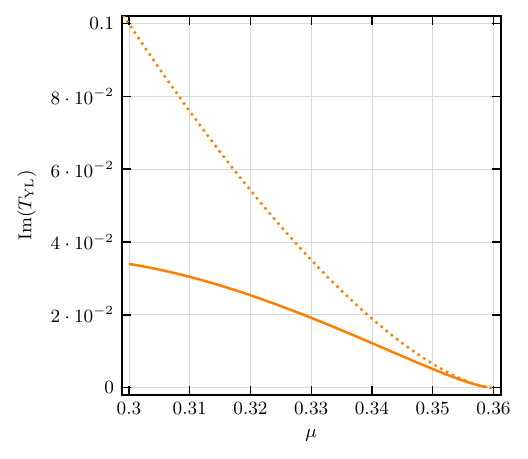}    
    \caption{Imaginary part of the Yang--Lee edge singularity as a function of chemical potential in the RMM, shown near small $\mu$ (left panel) and near the critical-point chemical potential (right panel).  The dotted curves show the corresponding asymptotic scaling forms.  At small $\mu$, the singularity initially moves away from the real axis; at larger $\mu$, the imaginary part turns over and eventually vanishes at the critical point.}
    \label{fig:ImT-rmm}
\end{figure}

Near the critical point, the singularity follows the universal scaling form derived in Sec.~\ref{sec:critical-scaling}.  In the present mean-field model, the leading critical behavior can be written as a Puiseux expansion,
\begin{equation}
T_{\rm YLE}(\mu)\approx T_c+ s(\mu-\mu_c)\pm i c s^{\Delta_{\rm MF}+1}|\mu-\mu_c|^{\Delta_{\rm MF}}
\end{equation}
where the coefficients $s$ and $c$ are 
\begin{eqnarray}
    s\equiv-\left.\frac{\partial_\mu\partial_\phi \Omega(\phi;T,\mu,\phi)}{\partial_T\partial_\phi  \Omega(\phi;T,\mu,\phi)}\right|_{\phi=\phi^*}
\end{eqnarray}
and
\begin{equation}
    c\equiv \left.\frac{2^{\Delta_{\rm MF}}}{3}\left|\frac{\partial^3 
    \Omega}{\partial\mu\partial\phi^2}\frac{\partial^2  \Omega}{\partial T\partial\phi}-\frac{\partial^3  \Omega}{\partial T\partial\phi^2}\frac{\partial^2  \Omega}{\partial \mu\partial\phi}\right|^{\Delta_{\rm MF}}\left(\frac{\partial^2 \Omega}{\partial_\mu\partial_\phi}\right)^{-\Delta_{\rm MF}-1}
    \left(\frac{\partial^4 \Omega}{\partial\phi^4}\right)^{\Delta_{\rm MF}-1}
    \right|_{\phi=\phi^*}
\end{equation}
respectively. 
The product of the mean-field critical exponents is $\Delta_{\rm MF} = \beta \delta = 3/2$.  The dashed curves in Fig.~\ref{fig:T-rmm} show this critical scaling prediction.  Comparison with the exact $T_{\rm YLE}(\mu)$ demonstrates that the critical region in this model is rather narrow. 

At small chemical potential, the same singularity admits an analytic expansion
    \begin{equation}
        T_{\rm YLE}(\mu)\approx T_0(m)+T_2(m) \mu^2+{\cal O}(\mu^4)
    \end{equation}
    with
    \begin{align}
        T_0(m)&\equiv\frac{e^{-i\theta}}{2\sqrt{3}\kappa}\left[
        -4e^{2i\theta}\kappa^{2/3}(2m^2-3)+e^{i\theta}\kappa^{1/3}(-16m^3+\kappa)+4m(-4m^3+\kappa(m))
        \right]^{1/2} 
        \\
        T_2(m) &\equiv-\frac{e^{2i\theta}\kappa^{2/3}(32m^5+(9-4m^2)\kappa)
        +72 e^{i\theta} \kappa^{4/3}m -128m^7+4m^2(4m^2+63)\kappa
    }{6T_0(m)(2m+e^{i\theta}\kappa^{1/3})^2\kappa}
    \end{align}
   where $\theta\equiv\pi/3$ and $\kappa(m)\equiv 2m(4m^2+27-3\sqrt{3}\sqrt{8m^2+27})$.
This approximation is shown in Fig.~\ref{fig:T-rmm} by dotted curves.  It gives a good description of the full result for all masses considered and for chemical potentials below $\mu\simeq0.25$.  As shown in Sec.~\ref{sec:critical-scaling}, the imaginary part of $T_{\rm YLE}$ must increase with $\mu^2$ at small $\mu$, while the real part decreases when the corresponding scaling combination has the sign realized in the model.  The left panel of Fig.~\ref{fig:ImT-rmm} displays this initial behavior, and the right panel shows the much narrower critical scaling region near the critical point.

The imaginary part of the trajectory provides a particularly clear diagnostic of two scaling regimes.  At small $\mu$, $\operatorname{Im}T_{\rm YLE}$ initially grows as $\mu^2$, as predicted by chiral scaling.  Closer to the critical point, the trend reverses and $\operatorname{Im}T_{\rm YLE}$ decreases to zero with the Ising exponent $\Delta_{\rm MF}$.  This turnover is less transparent when the same edge is represented in the complex chemical-potential plane (c.f. e.g. Ref.~\cite{Mukherjee:2021tyg}).

\begin{figure}[h]
    \centering
\includegraphics[width=0.35\linewidth]{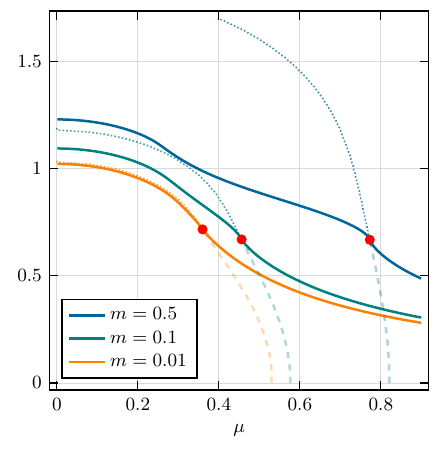}  
    \caption{RMM phase diagram for several values of $m$.  Solid curves show the real part of the leading complex-temperature singularity, dashed curves show the first-order transition line, and dotted curves show the maximum of the chiral susceptibility, a proxy for the chiral crossover transition.  Critical points are marked by red points.}
    \label{fig:pd-rmm}
\end{figure}

Figure~\ref{fig:pd-rmm} shows Re $T_{\rm YLE}$ as a function of the chemical potential and compares it to the location of the chiral susceptibility peak. We observe that, in the vicinity of the chiral limit, the two curves nearly coincide as expected. As the quark mass increases, however, a significant separation develops between them. This difference can be attributed to the nontrivial structure of the chiral susceptibility contours in the complex-temperature plane.

\subsection{Quark--meson model}
\label{sec:qm}

The quark--meson (QM) model provides a second test of the same analytic structure in a setting with a conventional order parameter and a tunable location for the critical point.  It also contains ``thermal'' singularities at values of $T$ where the Fermi--Dirac denominator vanishes at zero momentum~\cite{Skokov:2010uc,Skokov:2024fac}:
\begin{equation}
    \label{Eq:invFD}
    \exp \frac{m-\mu}{T} + 1 = 0; \quad \leadsto T = \pm i  \frac{m-\mu }{ \pi}.     
\end{equation}
These thermal singularities lie on the imaginary-$T$ axis and do not interfere with the critical singularities tracked below. This separation is an advantage of the complex-$T$ representation: in the complex chemical-potential plane, analogous thermal singularities can complicate the reconstruction of the critical analytic structure.

We use the mean-field thermodynamic potential
\begin{align}
    \Omega = \Omega_q(T,\mu, m) + \Omega_{q, {\rm vacuum} }(m)  +   \frac{\lambda}{4} (m^2-v^2)^2 - h m\,,     
\end{align}
where $m = g \sigma$ and the Yukawa coupling has been absorbed into the definition of the remaining parameters.  {We} take $\lambda = 1/5$ and $v=335$~MeV, which gives $T_c\simeq150$~MeV in the chiral limit.  The vacuum contribution requires regularization~\cite{Skokov:2010sf}; using dimensional regularization we write
\begin{align}
     \Omega_{q, {\rm vacuum} }(m)   =    -2 N_c N_f \int \frac{d^3 p }{(2\pi)^3} \sqrt{p^2+m^2} =   - \frac{N_c N_f}{8 \pi ^2} m^4 \ln \frac{m}{M}, 
\end{align}
where $M$ is an arbitrary normalization scale (we absorb all finite contributions into its definition).  

\begin{figure}[htb]
    \centering
    \includegraphics[width=0.45\linewidth]{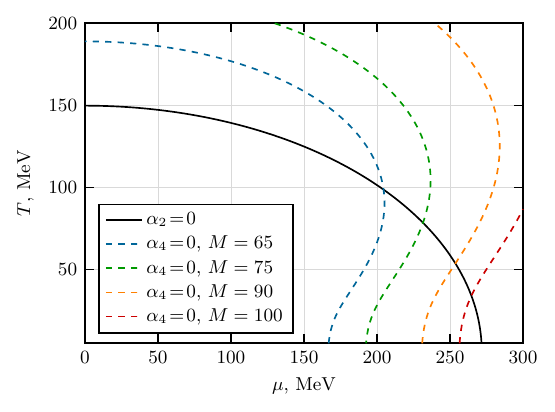}
    \caption{Chiral-limit structure of the QM model.  The curves show the zeroes of $\alpha_2$ and $\alpha_4$; their crossing defines the tricritical point.  The position of this point depends strongly on the vacuum scale $M$, allowing the nonzero-$h$ critical-point trajectory to be scanned over a range of slopes.}
    \label{fig:QMcrossing}
\end{figure}

To keep the relation with chiral scaling transparent while avoiding an exact second-order transition on the real axis, we work near the chiral limit and take $h=20~\mathrm{MeV}^3$. We expand the model's thermodynamic potential into power series in $m^2$: 
\begin{align}
    \Omega = \sum \alpha_{2n} m^{2n} - h m\, .
\end{align}
The expansion of the quark contribution $\Omega_q$ is given in App.~\ref{sec:Omegaq}. The non-analytic term $m^4 \log m^2$ cancels between $\Omega_q$ and the vacuum term, as was shown in Ref.~\cite{Skokov:2010sf}. 
Figure~\ref{fig:QMcrossing} shows the chiral-limit lines $\alpha_2=0$ and $\alpha_4=0$.  Their crossing defines the tricritical point, whose position depends strongly on the vacuum scale $M$.  At nonzero $h$, this dependence is inherited by the nearby critical point and changes the slope of the critical trajectory.  

\begin{figure}[htb]
    \centering
    \includegraphics[width=0.39\linewidth]{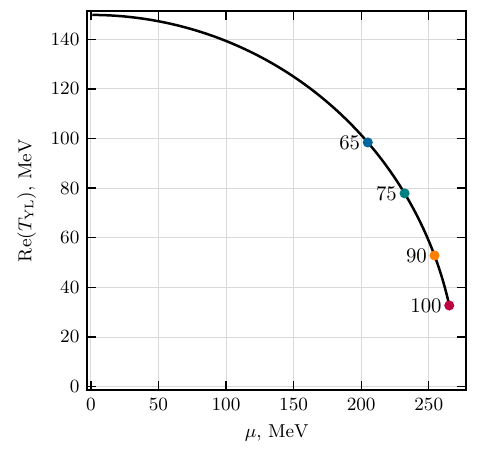}
    \includegraphics[width=0.37\linewidth]{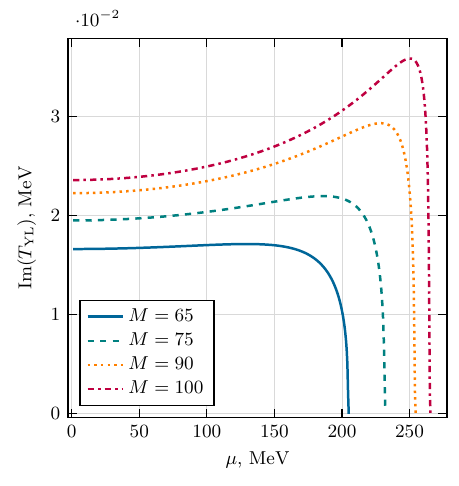}
    \caption{Real and imaginary parts of the leading YLE singularity in the complex $T$ plane as functions of chemical potential in the QM model.  For the real part, only the termination points at the critical point are shown because the trajectories nearly overlap.}
    \label{fig:QMreim}
\end{figure}

\begin{figure}[htb]
    \centering
    \includegraphics[width=0.49\linewidth]{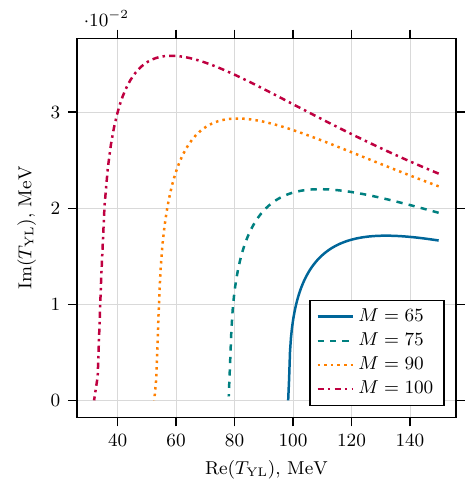}
    \caption{Trajectory of the leading Yang--Lee edge singularity in the complex $T$ plane in the quark--meson model for several values of the vacuum scale $M$.  Changing $M$ changes the slope of the critical line and therefore the approach of the edge singularity to the critical point.}
    \label{fig:QMphase}
\end{figure}

The corresponding YLE trajectories are shown in Figs.~\ref{fig:QMreim} and \ref{fig:QMphase}. The imaginary part of the complex-temperature singularity is particularly sensitive to the geometry of the critical line. Depending on the value of $M$, the singularity may move farther away from the real axis before bending back toward the critical point. As in the RMM, we observe a clear separation between two regimes in Im$\,T_{\rm YLE}$. At small $\mu$, the imaginary part increases with increasing chemical potential, while in the vicinity of the critical point it decreases. The two regimes are separated by a maximum of Im$\,T_{\rm YLE}$, whose prominence is controlled by the slope of the phase-transition line near the critical point. The same change of regimes is clearly observed in the full complex $T$ plane, see Fig.~\ref{fig:QMphase}.

\section{Critical scaling and its verification}

Near a critical point, the location of the Yang--Lee edge singularity is universally determined by the complex scaling variable $z_{\rm YLE}=t/h^{1/\Delta}$, where, in general,  the scaling fields $t$ and $h$ are functions of $T$ and $\mu$. This scaling relation holds throughout the critical regime, independent of whether the analytic continuation is carried out in the temperature or chemical-potential direction. In the following subsections, we exploit this universality to analyze the Yang--Lee edge singularities associated with the chiral phase transition and the QCD critical point.

\subsection{Chiral phase transition}

The relation between complex-$T$ and complex-$\mu$ singularities is especially transparent in the chiral scaling regime.  At small chemical potential, the reduced temperature can be written as
\begin{align}
\label{eq:t}
t/t_0 = \frac{T- T_c}{T_c} + \kappa_2 \left(\frac{\mu}{T}\right)^2 +  \kappa_4 \left(\frac{\mu}{T}\right)^4 + \ldots  \approx \frac{T- T_c}{T_c} + \kappa_2 \left(\frac{\mu}{T}\right)^2, 
\end{align}
and the symmetry-breaking field is $h=m_{u,d}/m_s$.  We neglect $\kappa_4$, whose lattice-QCD value is consistent with zero.  The location of the YLE singularity, $z_{\rm YLE}=t/h^{1/\Delta}$, is universal.  Therefore, within the scaling regime, complexifying $T$ or complexifying $\mu$ must give the same value of $z_{\rm YLE}$ and hence the same value of $t$:
\begin{equation}
   t _{\rm YLE} / t_0= \underbrace{T_{\rm YLE}/T_c-1 + \kappa_2 \left(\frac{\mu}{T_{\rm YLE}}\right)^2}_{\text{complexification of } T} = \underbrace{T/T_c - 1  + \kappa_2 \left(\frac{\mu_{\rm YLE}(T)}{T}\right)^2}_{\text{complexification of } \mu} \,. 
\end{equation}
where 
\begin{equation}
    t_{\rm YLE} \equiv z_{\rm YLE} h^{1/\Delta}
    \label{eq:tYLE}
\end{equation} 
is a complex number independent of $T$ and $\mu$. 
At $\mu=0$, this equality simplifies to
\begin{equation}
    \label{eq:Tyle}
    T_{\rm YLE}/T_c - T/T_c =  \kappa_2 \left(\frac{\mu_{\rm YLE}(T)}{T_c}\right)^2 \,.
\end{equation}

At the temperature $T = T^\star \equiv \mathrm{Re}\, T_{\rm YLE}(\mu=0)$, which isolates the real part of the complex-temperature edge singularity, Eq.~\eqref{eq:Tyle} gives
\begin{equation}
    i\, \mathrm{Im}\, T_{\rm YLE}/T_c =  \kappa_2 \left(\frac{\mu_{\rm YLE}(T^\star)}{ {\rm Re}\, T_{\rm YLE}}\right)^2 \,.
\end{equation}

Thus the square of the corresponding complex chemical potential, $(\mu^\star)^2 = \mu_{\rm YLE}^2(T^\star)$, is purely imaginary.  For the branch in the first quadrant this implies
\begin{equation}
    \mathrm{Re}\, \mu^\star = \mathrm{Im}\, \mu^\star \,,
\end{equation}
and fixes its magnitude through
\begin{equation}
    \mathrm{Im}\, T_{\rm YLE}/T_c =  2 \kappa_2 \left(\frac{\mathrm{Re}\, \mu^\star}{ {\rm Re}\, T_{\rm YLE}}\right)^2 \,.
\end{equation}

\begin{figure}[htb]
\includegraphics[width=0.45\textwidth]{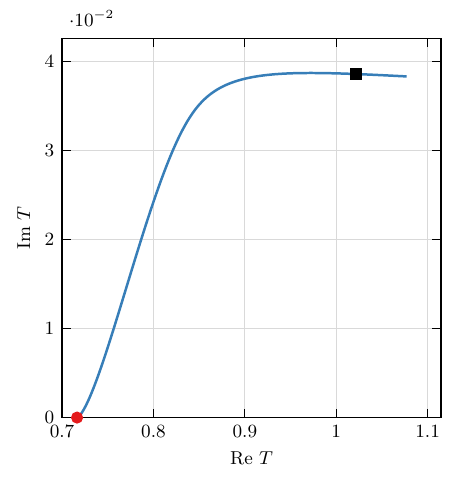}
    \hfill
\includegraphics[width=0.48\textwidth]{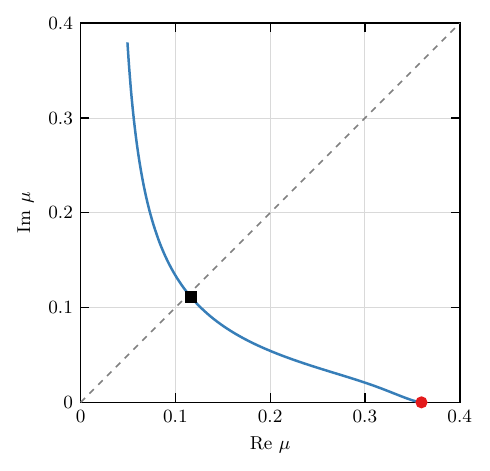}
    \caption{\label{Fig:thocomplexplanes} YLE singularity trajectories in two complementary complex planes in the RMM for $m=0.01$.  Left panel  shows the trajectory in the complex $T$ plane, $\{\text{Re }T_{\rm YLE}(\mu),\text{Im }T_{\rm YLE}(\mu)\}$, with $\mu$ ranging from imaginary values through $\mu=0$ (black square) to the critical point.  Right panel  shows the trajectory in the complex $\mu$ plane, $\{\text{Re }\mu_{\rm YLE}(T),\text{Im }\mu_{\rm YLE}(T)\}$, with $T$ ranging from $T_c$ to $1.4$.  The black square in right panel marks $T=\text{Re }T_{\rm YLE}(\mu=0)$, and red points mark the critical point.}
\end{figure}

This behavior is illustrated in the RMM in Fig.~\ref{Fig:thocomplexplanes}.  The black square in panel (a) denotes $T_{\rm YLE}(\mu=0)$, while the black square in panel (b) shows the corresponding $\mu^\star$.  In this model, $\mathrm{Re}\,\mu^\star$ is only approximately equal to $\mathrm{Im}\,\mu^\star$ because the exact scaling variable receives corrections to the quadratic approximation $t \approx T/T_c-1+\kappa_2(\mu/T)^2$.

The same relation can be compared with lattice-QCD information, assuming that the scaling variable remains useful at the physical pion mass.  In Ref.~\cite{Shah:2026pue}, extracting $T^\star$ from the condition $\mathrm{Re}\,\mu=\mathrm{Im}\,\mu$ gives $T^\star \approx 150$~MeV, close to the value of $\mathrm{Re}\,T_{\rm YLE}$ we find in Sec.~\ref{sec:lattice}.  The corresponding estimate $2\kappa_2T_c\left(\mathrm{Re}\,\mu^\star/{\rm Re}\,T_{\rm YLE}\right)^2\approx10$~MeV is also consistent with our result, see Sec.~\ref{sec:lattice}.  This agreement supports the use of Eq.~\eqref{eq:t} as an effective scaling variable near $T^\star$.
It is worth noting, however, that the data from Refs.~\cite{Clarke:2024ugt,Basar:2023nkp} do not appear to support a solution for $T^\star$ satisfying the condition $\mathrm{Re}\, \mu (T^\star) = \mathrm{Im}\, \mu (T^\star)$.

At nonzero but small chemical potential, Eq.~\eqref{eq:t} predicts a characteristic initial motion of the complex-temperature singularity.  Consider
\begin{align}
\frac{T_{\rm YLE}- T_c}{T_c} + \kappa_2 \left(\frac{\mu}{T_{\rm YLE}}\right)^2 = \frac{t_{\rm YLE}}{t_0} = \hat{t}
\end{align}
where $t_{\rm YLE}$ is a number independent of $T$ and $\mu$, defined in Eq. \eqref{eq:tYLE}.  Solving this equation for $T$ perturbatively in $\mu$ gives
\begin{align}
\frac{T_{\rm YLE}}{T_c}  = 1 + \hat t -  \kappa_2 \left(\frac{\mu}{T_{c}}\right)^2 \frac{(1+\hat t^*)^2}{|1+\hat t|^4} 
\end{align}
or separating real and imaginary parts 
\begin{align}
\frac{{\rm Re }\, T_{\rm YLE}}{T_c}  &= 1 + {\rm Re}\, \hat t -  \kappa_2 \left(\frac{\mu}{T_{c}}\right)^2 \frac{(1+ {\rm Re}\, \hat t)^2 - ({\rm Im}\, \hat t)^2}{|1+\hat t|^4} ,   \\ 
\frac{{\rm Im }\, T_{\rm YLE}}{T_c}  &= {\rm Im}\, \hat t  \left(1 +  2 \kappa_2 \left(\frac{\mu}{T_{c}}\right)^2 \frac{1+ {\rm Re}\, \hat t}{|1+\hat t|^4}\right)   \,.
\end{align}
Thus ${\rm Im}\,T_{\rm YLE}$ increases with $\mu^2$ at small $\mu$, while the real part can increase or decrease depending on the sign of $(1+{\rm Re}\,\hat t)^2-({\rm Im}\,\hat t)^2$.  In the lattice-QCD and model cases considered here this combination is positive, so ${\rm Re}\,T_{\rm YLE}$ decreases with $\mu$.

For the YLE singularity in the complex plane we have 
\begin{align}
   \frac{{\rm Re}\, \mu_{\rm YLE}}{T_c} &=  ({\rm Re} \,\hat t+1) \sqrt{\frac{\text{Im } \hat t}{2 \kappa }}-\frac{
   (\text{Re } \hat t +1  -2 \text{Im } \hat t)}{2  \sqrt{ 2\kappa  \text{Im } \hat t}} \left( \frac{T}{T_c} -1 - \text{Re } \hat t \right)+\ldots,\\
   \frac{{\rm Im}\, \mu_{\rm YLE}}{T_c} &=  ({\rm Re} \,\hat t+1) \sqrt{\frac{\text{Im } \hat t}{2 \kappa }}+\frac{
   (\text{Re } \hat t +1  +2 \text{Im } \hat t)}{2  \sqrt{ 2\kappa  \text{Im } \hat t}} \left( \frac{T}{T_c} -1 - \text{Re } \hat t \right)+\ldots\,.
\end{align}

These expressions show that ${\rm Im}\,\mu_{\rm YLE}$ increases with $T$, while the behavior of ${\rm Re}\,\mu_{\rm YLE}$ is controlled by the sign of $\text{Re}\,\hat t+1-2\text{Im}\,\hat t$.

\subsection{Critical scaling near a critical point}
\label{sec:critical-scaling}

Near the QCD critical point $(T_c, \mu_c)$, the equation of state is mapped
to Ising variables via the linear relations
\begin{align}
  r(T,\mu) &= r_T\,(T - T_c) - r_\mu\,(\mu_c - \mu),\\
  h(T,\mu) &= h_T\,(T - T_c) - h_\mu\,(\mu_c - \mu),
\end{align}
where $r$ and $h$ are the reduced temperature and magnetic field of the Ising
model, and $r_T, r_\mu, h_T, h_\mu$ are mapping coefficients.

At the critical point, scaling implies
\begin{equation}
  \label{eq:EOS}
  z_{\rm YLE}^{\Delta}\, h = r^{\Delta},
\end{equation}
where $\Delta = \beta \delta  > 1$ is a universal critical exponent and $z_{\rm YLE}$ is the universal position of the Yang--Lee edge singularity {in the three-dimensional $Z(2)$ universality class}. It is convenient to introduce $w_{\rm YLE} = 1/z^\Delta_{\rm YLE}$. 

Substituting the mapping and solving this equation perturbatively in $\mu_c-\mu$ we obtain a Puiseux series: 
\begin{equation}
  \label{eq:result}
  {
  T_{\rm YLE}(\mu) = T_c
    + \frac{h_\mu}{h_T}\,(\mu_c-\mu)
    + \frac{d^{\Delta} \,w_{\rm YLE}}{h_T^{\Delta+1}}\,(\mu_c-\mu)^{\Delta}
    + \cdots
  }
\end{equation}
where
\begin{equation}
  d = r_T h_\mu  - r_\mu h_T
\end{equation}
is the determinant of the mapping matrix. 

It is convenient to introduce 
\begin{align}
 s & = -\frac{h_\mu}{h_T}, \\ 
 c &= \frac{d^\Delta}{h_\mu^{\Delta+1}}.  
\end{align}
The parameter $s$ corresponds to the slope of the $r$ axis in the $T-\mu$ plane:
\begin{align}
    \tan \alpha = s\,.
\end{align}
In these variables we have 
\begin{equation}
  \label{eq:result-s}
  {
  T_{\rm YLE}(\mu) = T_c
    + s\,(\mu_c-\mu)
    + c \,w_{\rm YLE}s^{\Delta+1}\,(\mu_c-\mu)^{\Delta}
    + \cdots\,.
  }
\end{equation}
Taking into account that $z_{\rm YLE}^\Delta = \pm i |z_{\rm YLE}|^\Delta$, to the leading order, from Eq. \eqref{eq:result-s}  we obtain 
\begin{align}
    \label{eq:Re_Im_T}
    {\rm Re}\, T_{\rm YLE} &\approx  T_c 
              + s\,(\mu_c-\mu) \,,\\           
    {\rm Im}\, T_{\rm YLE} & \approx  \pm  c |w_{\rm YLE}| s^{\Delta+1}\,(\mu_c-\mu)^{\Delta}\,.  
\end{align}
This shows that the real part of the Yang--Lee edge temperature is a linear function of $\mu_c - \mu$, while the imaginary part of the temperature is proportional to $(\mu_c-\mu)^\Delta$.  

It is instructive to compare this to YLE scaling in the complex $\mu$-plane. 
Inverting Eq. \eqref{eq:result-s}, we can express the YL trajectory in complex-$\mu$ plane as a function of of real $T$
\begin{align}
    \mu_c - \mu_{\rm YLE} = \frac{1}{s} (T-T_c) - c \, w_{\rm YLE}\, (T-T_c)^\Delta + \cdots\,.
\end{align}
Thus 
\begin{align}
    \label{eq:Re_Im_mu}
    {\rm Re}\, \mu_{\rm YLE} &\approx  \mu_c 
              - \frac{1}{s}\,(T-T_c)\,, \\
    {\rm Im}\, \mu_{\rm YLE} &\approx \pm c |w_{\rm YLE}| (T-T_c)^\Delta\,. 
\end{align}
If the critical point is located by analyzing the complex-$\mu$ YLE trajectory using Eq. \eqref{eq:Re_Im_mu}, as in Refs.~\cite{Basar:2023nkp,Clarke:2024ugt}, the fit determines $T_c$, $\mu_c$, $s$, and $c|w_{\rm YLE}|$. The result can then be cross-validated by tracing the YLE trajectory in the complex $T$ plane using Eqs.~\eqref{eq:Re_Im_T}, since the same parameters determine both sets of equations.  This comparison can help distinguish genuine critical scaling from artifacts of analytic continuation or extrapolation.

Note that the expansion given in Eq. \eqref{eq:result} is with respect to the dimensionless parameter $\epsilon$: 
\begin{align}
    \epsilon = \frac{r_T \,w_{\rm YLE}  }{h_T^\Delta } \left[d (\mu_c-\mu)\right]^{\Delta-1}\,.
\end{align}
The expansion breaks down when this parameter becomes too large.  The critical value is
\begin{equation}
  \label{eq:ec}
  \epsilon_c = \frac{(\Delta-1)^{\Delta-1}}{\Delta^{\Delta}}.
\end{equation}
For {the three-dimensional Ising universality class}, 
$\Delta \approx 1.56$ and thus $\epsilon_c\approx 0.36$. 
For the mean-field value $\Delta=3/2$, Eq.~\eqref{eq:ec} gives $\epsilon_c=2/(3\sqrt{3})\simeq0.385$.  Comparing the truncated Puiseux expansion with the full model solution therefore provides a quantitative estimate of the range in $\mu$ over which the critical expansion is reliable.

As in the chiral scaling discussion, the critical scaling relation must hold regardless of whether one complexifies $T$ or $\mu$.  Recalling the linear mapping variables,
\begin{align}
  r(T,\mu) &= r_T\,(T - T_c) - r_\mu\,(\mu_c - \mu) \,, \\
  h(T,\mu) &= h_T\,(T - T_c) - h_\mu\,(\mu_c - \mu) \,,
\end{align}
and using the fact that the edge is located at $r^\Delta/h=z^\Delta_{\rm YLE}=\pm i\times\text{const}$ in the scaling regime, we obtain
\begin{equation}
    \underbrace{\frac{\left[r_T\,(T_{\rm YLE} - T_c) - r_\mu\,(\mu_c - \mu)\right]^\Delta }{h_T\,(T_{\rm YLE} - T_c) - h_\mu\,(\mu_c - \mu)}}_{\text{complexification of } T \text{ (parameter } \mu)} =  
    \underbrace{\frac{\left[r_T\,(T - T_c) - r_\mu\,(\mu_c - \mu_{\rm YLE})\right]^\Delta }{h_T\,(T - T_c) - h_\mu\,(\mu_c - \mu_{\rm YLE})}}_{\text{complexification of } \mu \text{ (parameter } T)} \,.
\end{equation}
This relation provides a direct consistency check of whether two independently reconstructed trajectories are governed by the same scaling regime.

\section{Analysis of zero-\texorpdfstring{$\mu$}{mu} lattice-QCD data}
\label{sec:lattice}

The current continuum estimate for the chiral-limit transition temperature at zero chemical potential is $T_c=132^{+3}_{-6}$~MeV~\cite{HotQCD:2019xnw}.  At physical quark masses, the chiral transition is replaced by a crossover; the associated Yang--Lee edge is therefore expected to move away from the real axis and to have a real part above the chiral-limit value of $T_c$.

Other singularities may also appear in the complex-temperature plane. In a conventional hadron resonance gas, singularities arise from the zeros of the Fermi--Dirac distribution functions (see Eq.~\eqref{Eq:invFD}), which are located on the imaginary-temperature axis. At $\mu=0$, fermionic modes generate singularities at $T=\pm i m/(n\pi)$ with $n\in\mathbb{Z}$, while bosonic modes lead to singularities at $T=\pm i m/(2\pi n)$. At nonzero chemical potential, these locations are shifted according to $m\to m\pm\mu$. {At $\mu=0$, both these thermal singularities and the singularities associated with Roberge--Weiss periodicity lie on the imaginary-temperature axis.} Consequently, {within the search window $\operatorname{Re}T\gtrsim50$~MeV, they do not compete with the chiral YLE candidate identified below}.


For each lattice spacing we analyze the baryon-number susceptibility $\chi^B_2(T)$ as a function of the real temperature and extract a candidate nearest singularity in the complex $T$ plane using an iterated conformal--Pad\'e construction \cite{Costin:2020pcj,Basar:2021hdf,Basar:2021gyi}.  The analysis is performed independently for $N_\tau\in\{10,12,16,20\}$ and for the two tree-level Symanzik improvement choices, denoted tr0 and tr1~\footnote{We thank Szabolcs Borsányi for generously providing the lattice QCD data from Ref.~\cite{Borsanyi:2021sxv}.
}.  Each input ensemble contains the temperature values, the sample mean, the quoted error, and 48 jackknife samples; the pole search is repeated independently for each jackknife bin.

For a given jackknife sample, we start from a complex seed
\begin{equation}
T^{(0)}_{\rm fish}=150+20\eta_1+i(10+4\eta_2)~{\rm MeV},
\end{equation}
where $\eta_1$ and $\eta_2$ are standard normal random variables.  At each iteration we construct a conformal map adapted to the current estimate $T_{\rm fish}$.  Writing $T_F=|T_{\rm fish}|$ and $\theta=\arg T_{\rm fish}$, each real temperature point is mapped to $\zeta\in(0,1)$ through
\begin{equation}
T=4T_F\zeta
\left[\frac{\theta/\pi}{(1-\zeta)^2}\right]^{\theta/\pi}
\left[\frac{1-\theta/\pi}{(1+\zeta)^2}\right]^{1-\theta/\pi}.
\end{equation}
In the $\zeta$ variable we fit near-diagonal Pad\'e approximants to $\chi^B_2(\zeta)$ using the three orders
\begin{equation}
[6/6],\qquad [7/7],\qquad [8/8].
\end{equation}
For each order we perform 20 random starts per jackknife sample.  The Pad\'e coefficients are determined by a least-squares minimization of the residual on the mapped real-temperature data.

The zeros of the Pad\'e denominator are then mapped back to the complex $T$ plane by the inverse form of the same conformal map.  We retain only poles satisfying
\begin{equation}
120~{\rm MeV}\leq {\rm Re}\,T\leq 260~{\rm MeV},\qquad
0.5~{\rm MeV}\leq |{\rm Im}\,T|\leq 80~{\rm MeV},
\end{equation}
together with $|{\rm Re}\,\zeta|<200$ and a minimum separation $|\zeta_{\rm pole}-\zeta_{\rm zero}|>0.03$ from all Pad\'e numerator zeros, in order to suppress near-canceling pole--zero defects.  At each iteration\textcolor{blue}{,} the updated estimate is chosen as the filtered pole with the smallest $|{\rm Im}\,T|$.  If no admissible pole is found, the next trial point is shifted by $+3$~MeV in the real part and $+2$~MeV in the imaginary part.  The iteration stops when
\begin{equation}
\left|T^{(n+1)}_{\rm fish}-T^{(n)}_{\rm fish}\right|<0.5~{\rm MeV},
\end{equation}
or after 30 iterations.

Each successful start contributes one final pole candidate, chosen as the mapped pole closest to the converged estimate.  The candidates obtained from all starts and all three Pad\'e orders are clustered in the complex $T$ plane with a default clustering radius of 5~MeV.  A cluster is accepted only if it contains at least three successful runs and receives contributions from at least two Pad\'e orders.  Among accepted clusters, we select the one with the largest Pad\'e-order multiplicity, then the largest number of runs, then the smallest compactness, and finally the smallest $|{\rm Im}\,T|$.  The jackknife-level pole estimate is taken to be the cluster medoid and is symmetrized as
\begin{equation}
T^{\rm (JK)}_{\rm sing}\to {\rm Re}\,T^{\rm (JK)}_{\rm sing}
+i\left|{\rm Im}\,T^{\rm (JK)}_{\rm sing}\right|.
\end{equation}

For each accepted jackknife estimate, we assign a systematic spread from two sources: the spread of the cluster medoids obtained separately for different Pad\'e orders and the variation under changing the clustering radius from 5~MeV to 4~MeV or 6~MeV.  These contributions are added in quadrature.  The final central values for ${\rm Re}\,T_{\rm sing}$ and ${\rm Im}\,T_{\rm sing}$ are the means over accepted jackknife estimates.  The statistical error is computed as
\begin{equation}
\sigma_{\rm stat}=\sqrt{\frac{N_{\rm JK}-1}{N_{\rm JK}}
\sum_{k=1}^{N_{\rm JK}}(x_k-\bar x)^2},
\end{equation}
while the quoted systematic error is the root-mean-square of the per-jackknife systematic estimates.

\begin{figure}
    \centering
    \includegraphics[width=0.45\linewidth]{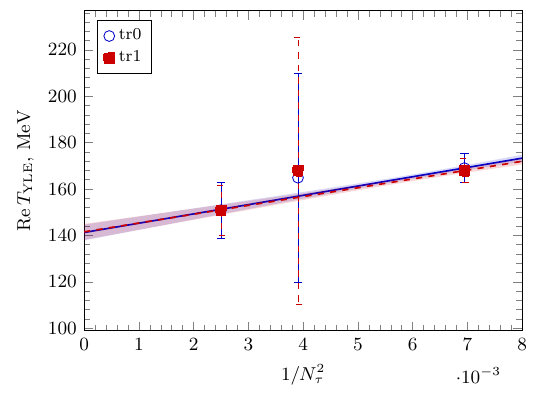}
    \includegraphics[width=0.45\linewidth]{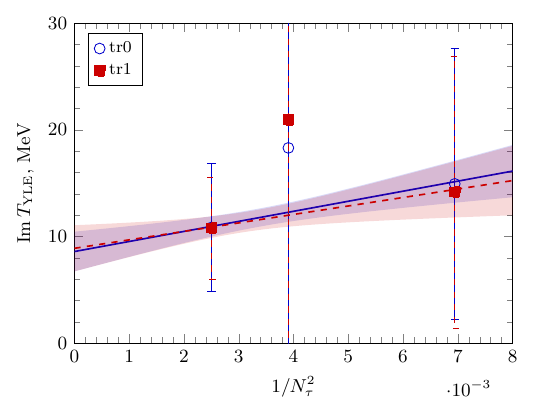}
    \caption{Continuum extrapolation of the real and imaginary parts of the nearest complex-temperature singularity extracted from lattice-QCD data at $\mu=0$.  The two sets of points correspond to the two tree-level improvement choices, while the intercept gives the continuum estimate.}
    \label{fig:lattice-continuum}
\end{figure}

The continuum extrapolation is performed independently for the real and imaginary parts using
\begin{equation}
T_{\rm sing}(x)=a+bx,
\qquad x=\frac{1}{N_\tau^2},
\end{equation}
with weights $w_i=1/\sigma_{{\rm stat},i}^2$.  The intercept $a$ is quoted as the continuum estimate.  The resulting extrapolation is shown in Fig.~\ref{fig:lattice-continuum}. The continuum extrapolated value is $T_{\rm YLE} = (141.23 \pm 3.44) \pm i (8.91 \pm 2.16)$ MeV and, as expected, is above the chiral phase transition temperature at $\mu=0$. Similar to RMM (see Fig.~\ref{fig:pd-rmm}), the value Re $T_{\rm YLE}$ is below the maximum of the chiral susceptibility $T_{\rm pc} \approx 150-157$ MeV, see Refs.~\cite{Aoki:2009sc,HotQCD:2018pds}. 
Nevertheless, the current result for $T_{\rm YLE}$  should be interpreted as a candidate singularity as its stability against the reconstruction choices is difficult to quantify reliably.

\section{Conclusions}
\label{sec:conclusions}

We have studied the analytic structure of the QCD phase diagram by treating the temperature as a complex variable.  The nearest Yang--Lee edge singularities in the complex $T$ plane provide information complementary to the more familiar complex-$\mu$ singularities.  Near a critical point, the two descriptions are governed by the same universal scaling equation of state and by the same nonuniversal map between QCD and Ising variables.

The random-matrix model shows two distinct regimes for the leading singularity in the complex-temperature plane.  At small chemical potential, the trajectory is analytic in $\mu^2$; in the model this expansion describes the exact result over a broad range of $\mu$.  Close to the critical point, the trajectory crosses over to the universal Puiseux form: the real part approaches the critical temperature linearly in $\mu_c-\mu$, while the imaginary part vanishes as $(\mu_c-\mu)^{\beta\delta}$.  The comparison with the exact solution shows that this critical scaling region is narrow.

The quark--meson model provides an independent realization of the same mechanism with a tunable critical-line geometry.  Varying the vacuum scale $M$ changes the location of the tricritical point in the chiral limit and, at nonzero symmetry-breaking field, changes the corresponding critical-point trajectory.  The resulting YLE trajectories show that the imaginary part of the complex-temperature singularity is sensitive to this geometry before it returns to the real axis at the critical point.

We also extracted the nearest complex-temperature singularity from lattice-QCD data at $\mu=0$ using an iterated conformal--Pad'e procedure. The continuum-extrapolated location, $T_{\rm YLE}\approx (141\pm i\,10)\,\mathrm{MeV}$, has a real part between the chiral-limit transition temperature and the chiral-susceptibility peak temperature at physical quark masses, while its nonzero imaginary part reflects the crossover nature of the transition.

Finally, the scaling relations derived here show how to test the same singularity structure by complexifying either $T$ or $\mu$.  Agreement between the two trajectories would support a critical-scaling interpretation of the extracted singularities, while disagreement would point to noncritical singularities or limitations of the analytic continuation.

\begin{acknowledgments}
We thank Szabolcs Bors\'anyi for useful discussions and for providing us with detailed LQCD data from Ref.~\cite{Borsanyi:2021sxv}. We are grateful to A. Kemper and S. Mukherjee for discussions.

This work is supported by the National Science Foundation CAREER Award PHY-2143149 (GB) and the U.S. Department of Energy, Office of Science, Office of Nuclear Physics through Contract {No.}~DE-SC0020081 (VS). 

We thank the Institute for Nuclear Theory at the University of Washington for its kind hospitality and stimulating research environment during the program INT-25-3a. This research was supported in part by the INT's U.S. Department of Energy grant No. DE-FG02-00ER41132.
\end{acknowledgments}

\appendix 

\section{Temperature-dependent part of the thermodynamic potential}
\label{sec:Omegaq}

Here we present the expansion of the temperature-dependent part of the thermodynamic potential
\begin{align}
    \frac{\Omega^{T}}{T^4} = \eta 
    \int \frac{d^(d-1) p}{(2\pi)^(d-1)} \left[ \ln \left(  1 -\eta e^{-\frac{\sqrt{p^2+m^2} -\mu }{T}  }   \right)  +  \ln\left(  1 - \eta e^{-\frac{\sqrt{p^2+m^2} +\mu }{T}  }   \right) \right]  
\end{align}
{as} a series in $x = m/T$.  In what follows we introduce $y=\mu/T$ to simplify the notation. 
Here $\eta = +1$ for bosons and $-1$ for fermions. To simplify notation we introduce $\nu=d/2$, where $d$ is the number of space-time dimensions.   
We have 
\begin{align}
    \frac{\Omega^{T}}{T^4}\, = - \sum_{i=1}^4 \mathcal{R}_i\,,   
\end{align}    
where 
\begin{align}
    \mathcal{R}_1 &= \pi^{-\nu}(\nu-1)! \sum_{l=0}^{\nu} \sum_{n=\delta_{l,\nu}}^{l} \frac{(-1)^{\nu+n+1}}{(2\nu-2l)! (l-n)! (2n)!} \notag \\
    &\quad \times (4^{n-l} - (1-\eta) 4^{n-\nu}) (2\pi)^{2\nu-2l} B_{2\nu-2l} (\nu-l+n-1)! x^{2l-2n} y^{2n}
\end{align}
Here $B_k$ are Bernoulli numbers.
The number of terms in this sum is finite.
The ``square-root'' term {is}
\begin{equation}
    \mathcal{R}_2 = \pi^{1-\nu} (\nu-1)! (-1)^\nu (1+\eta)  \frac{\nu!}{(2\nu)!} (x^2 - y^2)^{\nu-1/2}\,.
\end{equation}
This term does not contribute in the fermionic case. 
The ``logarithmic'' term {is}
\begin{equation}
    \mathcal{R}_3 = \pi^{-\nu} (\nu-1)! \frac{(-1)^\nu 2\eta}{4^\nu \nu!} x^{2\nu} \left[ \log\left(\frac{x}{4\pi}\right) + (1-\eta)\log 2 - \frac{1}{2} \sum_{k=1}^\nu \frac{1}{k} \right]\,.
\end{equation}
Finally, 
\begin{equation}
    \mathcal{R}_4 = \pi^{-\nu} (\nu-1)! \sum_{m=0}^{\infty} C_{2\nu+2m}(\nu, \eta, y) x^{2\nu+2m}
\end{equation}
where the coefficients are defined through the polygamma function $\psi^{(2m)}$:
\begin{align}
    C_{2\nu+2m}(\nu, \eta, y) &= \frac{(-1)^{\nu+m+1}}{m! (\nu+m)!} \frac{1}{2^{4m+2\nu} \pi^{2m}} \Big[ \psi^{(2m)}\left(1 - \frac{iy}{2\pi}\right) + \psi^{(2m)}\left(1 + \frac{iy}{2\pi}\right) \notag \\
    &\quad + 4^m (\eta - 1) \left( \psi^{(2m)}\left(1 - \frac{iy}{\pi}\right) + \psi^{(2m)}\left(1 + \frac{iy}{\pi}\right) \right) \Big]\,.
\end{align}

Series expansion of these functions in power series in $y$ reproduces the high-temperature expansion series of Ref.~\cite{Landsman:1986uw}.

\section{Special case of critical point mapping \texorpdfstring{$h_T = 0$}{ht = 0}}

Section~\ref{sec:critical-scaling} discussed the convergence of the Puiseux expansion and the constraints it places on the range of $\mu$. 

The convergence properties are especially important when $h_T = 0$, as the limits $\mu_c-\mu \to 0$ and $h_T\to 0 $ do not commute.

When $h_T = 0$ the $h=0$ axis becomes vertical in the $(T,\mu)$ plane
(slope $dT/d\mu = -h_\mu/h_T \to -\infty$), meaning $h$ is no longer mixed
with temperature and reduces to
\begin{equation}
  h = -h_\mu\,(\mu_c-\mu).
\end{equation}
The equation of state \eqref{eq:EOS} then becomes:
\begin{equation}
  z^{\Delta}\,(-h_\mu\,(\mu_c-\mu) ) = r(T,\mu)^{\Delta}
  \quad\Longrightarrow\quad
  r = z\,(-h_\mu)^{1/\Delta}\,(\mu_c-\mu)^{1/\Delta},
\end{equation}
and expanding $r = r_T(T-T_c) - r_\mu\,(\mu_c-\mu)$ gives
\begin{equation}
  \label{eq:ht0}
  {
  T(\mu)\Big|_{h_T=0} = T_c
    + \frac{z\,(-h_\mu)^{1/\Delta}}{r_T}\,(\mu_c-\mu)^{1/\Delta}
    + \frac{r_\mu}{r_T}\,(\mu_c-\mu). 
  }
\end{equation}
This is qualitatively different from the generic expansion \eqref{eq:result-s}.

Considering only the leading order, we have 
\begin{align}
    {\rm Re}\, T &\approx  T_c 
              + \frac{|z_{\rm YLE}| \cos(\frac{\pi}{2 \Delta}) (-h_\mu)^{1/\Delta}}{r_T}\,(\mu_c-\mu)^{1/\Delta} \\           
    {\rm Im}\, T & \approx  \pm  \frac{|z_{\rm YLE}| \sin(\frac{\pi}{2 \Delta}) (-h_\mu)^{1/\Delta}}{r_T}\,(\mu_c-\mu)^{1/\Delta}  \,.  
\end{align}

\bibliography{apssamp}

\end{document}